\documentclass[hyper]{JHEP3}

\input{epsf}
\usepackage{epsfig}
\usepackage{amssymb}
\usepackage{amsfonts}
\usepackage{amsbsy}
\usepackage[all]{xy}
\usepackage{amsmath}

\usepackage{amssymb,amscd}
\usepackage{mathrsfs}
\usepackage{amsmath,amsthm}
\def\be{ \begin{equation} }
\def\ee{ \end{equation}}

\def\Aut{{\rm Aut}}

\def\one{{\hbox{ 1\kern-.8mm l}}}
\def\CC {{\cal C}}
\def\CD {{\cal D}}

\def\CI {{\cal I}}

\def\CL {{\cal L}}

\def\CX {{\cal X}}

\def\CI {{{\cal I}}}

\def\CX {{\cal X}}

\def\IC{\mathbb{C}}

\def\IM{\mathbb{M}}

\def\IQ{\mathbb{Q}}
\def\IR{{\mathbb{R}}}

\def\IZ{{\mathbb{Z}}}

\def\rmk#1{\bigskip\noindent{\bf Remarks} }
\def\rk{{\rm Rank}}

\title{\boldmath Hyperk\"ahler Isometries Of K3 Surfaces}

\author{Anindya Banerjee and Gregory W. Moore \\
 NHETC and
$~~$Department of Physics and Astronomy, Rutgers University \\
$~~$126 Frelinghuysen Rd., Piscataway NJ 08855, USA\\
\\
{\tt ab1702@scarletmail.rutgers.edu, gmoore@physics.rutgers.edu } }

\abstract{We consider symmetries of K3 manifolds. Holomorphic symplectic
automorphisms of K3 surfaces have been classified, and observed to be
subgroups of the Mathieu group $M_{23}$. More recently, automorphisms of
K3 sigma models commuting with $SU(2)\times SU(2)$ $R$-symmetry have been
classified by Gaberdiel, Hohenegger, and Volpato. These groups are all
subgroups of the Conway group. We fill in a small gap
in the literature and classify the possible hyperk\"ahler isometry groups
of K3 manifolds. There is an explicit list of $40$ possible groups, all of which
are realized in the moduli space. The groups are all subgroups of $M_{23}$.\\ \today}
\begin{document}
\maketitle
\flushbottom

\section{Introduction And Conclusion}

The study of K3 manifolds has led to many fascinating
 developments in both mathematics and physics.
Among their many exceptional properties these manifolds
exhibit interesting discrete symmetry groups. This paper
fills a surprising gap in the literature on the symmetries
of K3 manifolds: We classify the hyperk\"ahler isometries
of such manifolds.

In the literature one will find slightly different definitions
of what is   meant by ``K3 manifold,'' and the distinctions
between the definitions will be very important to us here.
Throughout this paper
\footnote{With the exception of an example discussed in
Appendix \ref{app:InfiniteHoloSymp}.}
$X$ will be a \underline{smooth} compact oriented simply connected four-manifold.

We will assume that $X$ admits an integrable complex structure $J$
such that the canonical bundle $\cal{K}$ is holomorphically trivializable.
When a choice of such a $J$ is made we will refer to the pair $(X,J)$ as a \emph{K3 surface}.
A choice of holomorphic trivialization of  $\cal{K}$
defines the structure of a holomorphic symplectic
manifold. In this paper when we speak of $(X,J)$
one can assume that such a trivialization has been chosen.
It is unique up to overall scale, and the choice of
scale will not affect our considerations.
By a ``symmetry of a K3 surface'' one could mean a holomorphic symplectic
automorphism of $(X,J)$. That is, a holomorphic
diffeomorphism $f:(X,J) \to (X,J)$ that acts
trivially on the one-dimensional cohomology $H^{2,0}(X;\IC)$.
Some K3 surfaces $(X,J)$ have interesting groups of
holomorphic symplectic automorphisms and these have been
extensively studied in the literature
\cite{Hashi,Hoehn:2014ika,Kon,MASONREVIEW,Muk,Xiao}.
See \cite{HuyK3}, chapter 15 for a nice exposition of known results.

On the other hand, by a \emph{K3 manifold} we mean a pair
$(X,g)$ where $g$ is a smooth hyperk\"ahler metric on $X$,
that is, $(X,g)$ is a Riemannian manifold such that
there is a triplet of parallel integrable complex
structures $J_i$, $i=1,2,3$ satisfying the quaternionic relations
\be\label{Quaternion}
J_i J_j = - \delta_{ij} + \epsilon_{ijk} J_k ~ .
\ee
By a ``hyperk\"ahler isometry of $(X,g)$'' we mean
a diffeomorphism $f: X \to X$ which is an isometry
of $g$ such that $f^*J_i= J_i$. The aim of this paper
is to identify what groups can and do appear as groups
of hyperk\"ahler isometries for some K3 manifold $(X,g)$. When we speak of
\underline{the} hyperk\"ahler isometry group for some specific $(X,g)$ we always mean
the maximal group of such isometries. Otherwise we use the
term ``a group of hyperk\"ahler isometries of $(X,g)$.''  The  hyperk\"ahler
isometry group of $(X,g)$ is always a subgroup of the group of
all isometries of $(X,g)$. The isometry groups of K3 manifolds
are certainly of interest, but unfortunately we will not have much to say about them
in this paper aside from one remark in section \ref{sec:FutureDirections} below.

While hyperk\"ahler isometry groups are always finite groups
it is possible for $(X,J)$ to have
an infinite group of holomorphic symplectic automorphisms
(see appendix \ref{app:InfiniteHoloSymp} for an example).
In this paper we will always consider finite groups
of holomorphic symplectic automorphisms of K3 surfaces.
Thus, there are several lists of ``K3 symmetry groups''
we could compile:
\footnote{These are lists of \underline{isomorphism classes} of groups.}

\begin{enumerate}

\item $\CL_{HolSymp}$ is the set of finite groups $G$ such that $G$ acts as a group of symplectic automorphisms of
\underline{some} $K3$ surface $(X,J)$.  We will refer to this as the Xiao-Hashimoto list \cite{Hashi,Xiao}.

\item  $\CL_{HolSymp}^{max}$ is the subset of $\CL_{HolSymp}$ such that there is an $(X,J)$
such that $G$ is not a proper subgroup of any other finite group which also has a
symplectic action on that surface $(X,J)$. We will refer to this as Mukai's list \cite{Muk}.

\item $\CL_{HKIsom}$ is the list of groups $G$ that appear as the hyperk\"ahler isometry group of \underline{some} K3 manifold $(X,g)$.

\item $\CL_{IsomHK}$ is the list of groups which are maximal groups of isometries of some hyperk\"ahler $(X,g)$. It is obvious
that $\CL_{HKIsom} \subset \CL_{IsomHK}$, but as we just noted, we do not know much   about this last list of groups.

\end{enumerate}

As we will discuss in the next section, simple considerations show that
\begin{equation}\label{eq:Inclusions}
    \CL_{HolSymp}^{max} \subset \CL_{HKIsom} \subset \CL_{HolSymp}
\end{equation}

More nontrivially, we can relate these lists to another list of groups,
$\CL_{ConwayFix}$ which is the list subgroups of the Conway group $Co_0$ (considered up to conjugacy)
whose action on the Leech lattice $\Lambda$ has a nontrivial sublattice of fixed vectors. This list was compiled by
  H{\"o}hn and Mason and we refer to it as H{\"o}hn and Mason's list \cite{HM}. More precisely,
  $G\in \CL_{ConwayFix}$ if $\Lambda^G \not= \{ 0 \}$ and $G$ is not a proper subgroup of
  some other subgroup $H\subset Co_0$ with $\Lambda^H = \Lambda^G$. (See Appendix \ref{app:LatticeResults}
  for our notation for lattices.)

We will show that
\be
\CL_{HKIsom} \subset \CL_{ConwayFix}
\ee
and in fact
\be\label{eq:MainResult}
\CL_{HKIsom} = \CL_{HolSymp} \cap \CL_{ConwayFix}
\ee
and as a consequence we learn that each of the inclusions in \eqref{eq:Inclusions} is proper.
This is the main result of this paper: The list of possible hyperk\"ahler isometry groups
is given in Appendix \ref{app:HyperKList}.

Our work here was mostly motivated by physical questions about heterotic/M-theory duality
as well as questions related to Mathieu Moonshine. We discuss the relations to physics in
section \ref{sec:PhysicalImplications}. The techniques we use are lattice-theoretic, and
are closely related to the techniques used by \cite{Gaberdiel:2011fg,Kon,Huy,Xiao}.
The known results on symplectic automorphisms of K3 surfaces were reproduced using the
kind of techniques used here in \cite{Hoehn:2014ika}.

This paper was motivated by some questions raised in a project with J. Harvey
investigating Mathieu Moonshine \cite{Harvey:2017rko,Harvey:2017xdt,Harvey:2020jvu}.
That project is closely
related to the work of Gaberdiel, Hohenegger and Volpato \cite{Gaberdiel:2011fg}. Indeed, the GHV
theorem gives yet another list of groups associated to K3 manifolds, namely, the list of
automorphism groups of the two-dimensional sigma model with $(X,g,B)$ as a target manifold.
(Here $B$ is a flat ``$B$-field'' or ``$U(1)$ gerbe connection.'') By automorphism
group we mean an automorphism of the two-dimensional conformal field theory commuting
with the $(4,4)$ superconformal algebra.  The GHV list
$\CL_{SigAut}$ is also a sublist of $\CL_{ConwayFix}$ and contains $\CL_{HKIsom}$ as a
proper sublist - it is the list of ``classical'' or ``geometrical'' automorphisms
of K3 sigma models.
%
%The technical methods used in this paper rely on lattice theory
%and in particular we have closely followed the techniques used in
% the papers  \cite{Gaberdiel:2011fg} and  \cite{Huy}.

\section*{Acknowledgements}

We thank the referee for pointing out an error in our statement of Lemma 1 in v1 of this paper.
Fortunately, as we explain in the comments following
our revised version of Lemma 1, the remainder of the paper is unaltered.
We also thank  J.~ Harvey and  D.~ Morrison for discussions and collaboration
on related matters, as well as useful remarks related to the draft. We also thank G.~H{\"o}hn, D.~ Huybrechts, and
G.~Mason for useful comments and discussion.
 This work is supported by the US DOE   under grant
DOE-SC0010008 to Rutgers.

\section{Some Easy Relations Between The Lists}

As we have mentioned, our approach to understanding the symmetries of K3 manifolds
 is essentially lattice-theoretic. We explain this point first:

To $(X,g)$ we
 can associate an oriented, positive-definite 3-plane, $P$ in the quadratic real vector space $H^2(X;\IR)$
 of signature $(+^3, -^{19})$. Indeed, given a compatible complex structure for $(X,g)$
 we can define a K\"ahler form by $\omega(v,w) = g(Jv,w)$
for $v,w \in TX\otimes \IC$.  Then $P$ is the span of the classes of the three K\"ahler
forms $[\omega_i]$ associated to $J_i$, and there exists a natural orientation for this basis, where the K{\"a}hler classes are oriented so as to preserve (\ref{Quaternion}).

Conversely, given an oriented positive
3-plane $P \subset H^2(X;\IR)$ we may associate a hyperk\"ahler manifold $(X,g)$, unique
up to diffeomorphism and an overall choice of volume. To see this,
given  $P$ a compatible complex structure $J$
corresponds to a choice of two-dimensional subspace $S\subset P$.
$S$ is to be identified with the subspace spanned by the real and complex
components of the holomorphic symplectic form, while the orthogonal
complement to $S$ inside $P$ is a one dimensional space. One can choose any
nonzero vector in $S^\perp \cap P$ as a K\"ahler class. Thanks to Yau's theorem
there then exists a unique hyperk\"ahler metric $(X,g)$ compatible with $J$
and inducing that K\"ahler class. It follows that to $P$ we can associate
a HK manifold $(X,g)$ by choosing, say, unit volume.  Therefore, the moduli space
\footnote{actually, coarse moduli stack}
 of possible K3 manifolds $(X,g)$ (with unit volume)  may be
identified with the double coset
\be\label{eq:HK-ModSpace}
O^+_{\IZ}(H^2(X,\IZ)) \backslash O^+_{\IR}(H^2(X;\IR)) /SO(3) \times O(19) ~ .
\ee
On the left we quotient by an index two subgroup $O^+_{\IZ}(H^2(X,\IZ)) \subset \Aut(H^2(X,\IZ))$
that has an orientation-preserving action on the oriented positive 3-planes in $H^2(X;\IR)$.
By a result of Donaldson \cite{Don,HuyK3} this is the full group of (isotopy classes of) orientation-preserving
diffeomeorphisms of $X$.
Not all the metrics in \eqref{eq:HK-ModSpace} are smooth. Since we only consider
smooth manifolds in this paper we will always restrict attention to subspaces $P\subset H^2(X;\IR)$
so that the orthogonal space $P^\perp$ does not contain any vectors $\alpha$ with $\alpha^2=-2$.
Such vectors are called root vectors.

Next,    the lattice $H^2(X;\IZ)$ is an even unimodular integral lattice of signature $(3,19)$ and
consequently there is an isomorphism of lattices:
\be\label{eq:K3-Lattice}
H^2(X;\IZ) \cong  II^{3,19} \cong  E_8(-1)^{\oplus 2} \oplus U^{\oplus 3}.
\ee
where $U=II^{1,1}$ is the standard hyperbolic plane, spanned by $e,f$ with $e^2 = f^2 =0 $ and $e\cdot f = 1$.
In the text we will simply write $\Gamma$ for $II^{3,19}$ to simplify the notation.

Thanks to the above facts, the hyperk\"ahler isometry groups we seek are
precisely the groups $G\subset O^+(\Gamma)$ acting trivially on a
positive 3-plane $P\subset \Gamma^{3,19}\otimes \IR$ that contains no
root vectors in $P^\perp$. It is interesting to contrast that with the
formulation of an isometry. The latter would be a diffeomorphism $f$
such that the action of $f^*$ on $H^2(X;\IR)$ preserves the set $P$ but
might act as a nontrivial isometry of $P$. Thus, the set of such diffeomorphisms must form a discrete
subgroup of $O(3)$ and hence any isometry group of a hyperk{\"a}hler metric on a K3 manifold must be a finite group.

Now, any  hyperk\"ahler isometry group for $(X,g)$ acts trivially on the
corresponding 3-plane $P$ and hence acts trivially on any two-dimensional
subspace $S \subset P$. It is consequently a finite group of
holomorphic symplectic automorphisms for any compatible $J$.
It follows that $\CL_{HKIsom} \subset \CL_{HolSymp}$. A more substantive
claim is that $\CL_{HolSymp}^{max} \subset \CL_{HKIsom}$. To see this
suppose that $G$ acts as a maximal finite group of holomorphic symplectic
automorphisms of some $(X,J)$. The span of the real and imaginary parts
of a compatible holomorphic symplectic form is a positive 2-plane
$S \subset H^2(X;\IR)$ and $G$ acts trivially on $S$.
 Choose any K\"ahler class $\beta$ so that
$P = S \oplus \beta \IR$ is a positive 3-plane. In general $G$ will
not fix $\beta$, however, for $f\in G$, $f^*$ preserves the K\"ahler
cone, and the K\"ahler cone is convex so that
\be
\tilde \beta := \sum_{f\in G} f^*\beta
\ee
is an invariant K\"ahler class. Therefore, $G$ is a group of hyperk\"ahler
isometries of the K3 manifold $(X,g)$ associated to $\tilde P = S \oplus \tilde\beta \IR$.

\section{Relating Hyperk\"ahler Isometry Groups To Subgroups Of The Conway Group}

We now come to a useful pair of lemmas, where we relate groups of hyperk\"ahler isometries
to the Conway group $Co_0$ of automorphisms of the Leech lattice $\Lambda$. Our notation for
lattices and related concepts can be found in Appendix \ref{app:LatticeResults}.

As a preliminary, we  will repeatedly make use of the following elementary

\bigskip
\noindent
\textbf{Lemma 1}: Suppose $\iota: L\hookrightarrow M$ is a primitive embedding of
an even integral lattice $L$ into an even unimodular lattice $M$.
Suppose that a group $G$ acts as a group of automorphisms
of $L$ such that $L^G = \{0\}$ and the action of $G$ on the discriminant group $\mathcal{D}(L)$ is trivial.
 Then   $G$ acts as a group of automorphisms on $M$ so that
$\iota(L) = M_G = (M^G)^\perp$.

\bigskip
\bigskip
\noindent
\emph{Proof}: Let $N= \iota(L)^\perp$ where the orthogonal complement
is taken within $M$. Since the lattice $M$ is
self-dual  there is an isometric isomorphism of discriminant groups
\be
\psi: (\CD(L), q_L) \to  (\CD(N), q_N) ~ .
\ee
such that $\psi^*(q_N) = - q_L$   \cite{MirandaMorrison,Nik}. 
Using the discriminant group one can construct an isomorphic
copy of $M$ from the sublattice $\iota(L) \oplus N$ using
``glue vectors.'' To be precise, we consider the lattice
\be
\iota(L)^\vee \oplus N^\vee \subset M \otimes \IQ
\ee
and take the fiber product, i.e. the vectors
$\ell\oplus n \in \iota(L)^\vee \oplus N^\vee$ such that
\be
\psi([\ell]) = [n].
\ee
The fiber product is isomorphic to $M$ \cite{MirandaMorrison,Nik}. 
Now, if we posit that $G$ acts trivially on $N^{\vee}$,
we can always induce a $G$-action on $\iota(L)^{\vee}\oplus N^{\vee}$. Generically, such a
$G$ action will not preserve the sublattice $M$ setwise, so $G$ will not be an automorphism of $M$.
\footnote{As an example, provided by the referee, consider a primitive embedding of $L=\sqrt{N} \IZ$ 
into $II^{1,1}$.}
However, if $G$ also acts trivially on $\mathcal{D}(L)$, then for
any $x \in \iota(L)^{\vee}$, we have, $g\cdot x=x\text{ (mod }(L))$ and hence $g\cdot x-x \in \iota(L)$. 
Now, for any such $x\in \iota(L)^\vee$, there exists $y\in N^{\vee}$ such that $v=x\oplus y \in M$. Then, $g\cdot v-v=g\cdot(x\oplus y)-(x\oplus y)=(g\cdot x-x) \in M$. Thus, $G$ is an automorphism of $M$. 
Therefore, the natural $G$-action on $\iota(L)^\vee \oplus N^\vee$ with $G$ acting trivially on $N^\vee$ can be used to induce a $G$-action on the sublattice isomorphic to $M$, provided $G$ acts trivially on the discriminant group of $L$.
Since $L^G=\{0\}$ it follows that $M^G = N$ and hence $\iota(L) = M_G$.   $\spadesuit$

\bigskip
\noindent
\textbf{Remark}: We now amplify on the need for $G$ to act trivially on $\CD(L)$.  Consider 
 an even unimodular lattice $\Gamma$ such that a finite group $G$ acts as an automorphism group. Let $N\cong \Gamma^G$ be the sublattice of $\Gamma$ on which $G$ acts trivially. This implies that the induced $G$-action on $\mathcal{D}(N)$ is trivial. We can then construct $L$ as the orthogonal complement of $N$, and prove that the group $G$ will act trivially on the discriminant group of $L$.
We can demonstrate this using the argument in the proof above, but in reverse: 
If $x \in L^\vee$ then there is a $y\in N^\vee$ so that
$v= x \oplus y \in \Gamma$, because $\psi$ is an \underline{isomorphism}.
But then for every $g\in G$, we have  $g\cdot v - v \in \Gamma$ but also
$g\cdot v - v = g\cdot(x \oplus y) - (x \oplus y) = g \cdot x - x $. So $g\cdot x = x \text{ (mod } L)$,
so the $G$ action on $\CD(L)$ is trivial. \\

Over the next sections, we will often invoke the following chain of arguments: We begin with any even unimodular lattice $\Gamma$ and a finite group $G\subset \text{Aut}(\Gamma)$. We choose a sublattice $L\cong \Gamma_G$, and as we have seen,   $G$ acts trivially on $\mathcal{D}(L)$. 
Next we  embed $L$ primitively into a different even unimodular lattice $M$ ($\iota:L\hookrightarrow M$). The conditions of Lemma 1 
are satisfied and we can use it to extend the $G$-action from $\iota(L)$ to $M$. As an example of this procedure, our  next lemma tells us that suitable subgroups of $G \in \Aut(\Gamma)$, where $\Gamma$ is the K3 lattice,  can
be ``transferred'' to isomorphic groups $G\in \Aut(\Lambda)$, where $\Lambda$ is
the Leech lattice. The transfer works both ways.

\bigskip
\bigskip
\noindent
\textbf{Lemma 2}:

The following two conditions on a finite group $G$ are equivalent:

\begin{enumerate}

\item  $G$ is a subgroup of  $O^+(\Gamma)$ acting trivially on a positive 3-plane $P\subset \Gamma\otimes \mathbb{R}$ with no root vectors in $P^{\perp}$

\item  $G\subset Co_0$ and $G$ acts on the Leech lattice $\Lambda$ with an invariant sublattice $\Lambda^G$
of rank $R$ and discriminant form $q:\CD(\Lambda^G) \to \IQ/\IZ$ such that there exists an even integral lattice with
invariants
\footnote{See equation \eqref{eq:InvtsDef} for the definition of the invariants.}
 $\{ ( 3, R-5, -q)\}$. (In particular, $R\geq 5$.)

\end{enumerate}

We now prove  Lemma 2.
In the next section, we prove that Condition 1 implies Condition 2. In the subsequent section, we show that Condition 2 implies Condition 1. Together, this gives us a bijection.

\subsection{Condition 1 Implies Condition 2}

Recall that we denote the K3 lattice \eqref{eq:K3-Lattice} by $\Gamma$.
The main step will be to embed $\Gamma_G(-1)$ into the Leech lattice $\Lambda$.
Our strategy for doing this will be to use the formulation of the Leech lattices
as a subquotient of $II^{1,25}$, namely
\be
\Lambda(-1) \cong \textbf{n}^\perp/\textbf{n}\IZ
\ee
where $\textbf{n}\in II^{1,25}$ is the famous null vector $\textbf{n}=(70;0,1,2,\dots, 24)$.
Thus, our first step will be to try to embed $\Gamma_G$ into $II^{1,25}$, and then argue that there
exists and embedding which descends to an embedding into the subquotient.

Let us first check for the existence of an embedding of $\Gamma_G$ into $II^{1,25}$. To do this we use
 Nikulin's theorem 1.12.2 \cite{Nik} on the conditions for the existence of a primitive embedding
 of an even lattice into an even unimodular lattice. The full theorem is described in
 Appendix \ref{app:LatticeResults}. Since $\Gamma_G$ is negative definite and of rank $\leq 19$
 we need only check the sufficient condition \eqref{eq:NikIneq}. Written out in our case
 this becomes:
\be\label{eq:FirstCheck}
26 > \rk(\Gamma_G) + \ell(\Gamma_G) ~ .
\ee
We now express this in terms of $\Gamma^G$ using
$\rk(\Gamma_G) = 22- \rk(\Gamma^G)$ and
$ \ell(\Gamma_G) = \ell(\Gamma^G)$. The
second equation follows because $\Gamma^G$ and $\Gamma_G$
are orthogonal primitively embedded lattices in an
even unimodular lattice so that $\CD(\Gamma_G) \cong \CD(\Gamma^G)$.
Therefore \eqref{eq:FirstCheck}
is  equivalent to
\be\label{eq:FirstCheck2}
4  > \ell(\CD_{\Gamma^G})  - \rk(\Gamma^G)
\ee
Now we simply use \eqref{eq:RankEll}. Hence the required embedding exists.

Now we wish to show that we can take the embedded copy $\iota(\Gamma_G) \subset II^{1,25}$
to be in $\textbf{n}^\perp$, that the embedding remains an embedding after passing
to the quotient by $\textbf{n}\IZ$,   and that $G$ is a group
of automorphisms of the subquotient.

%
%On the other hand $\iota(\Gamma_G)$ is
%orthogonal to at least one root vector, namely the generator of $\iota(\CD_1(-1))$.
%

Using the techniques outlined in Remark 1, we can demonstrate that $G$ acts trivially on the discriminant group of $\Gamma_G$. Since $G$ acts nontrivially on $\Gamma_G$ with no fixed-point sublattice
we can then use Lemma 1 to define an action of $G$ on $II^{1,25}$ as a group
of automorphisms so that $\Pi:= \iota(\Gamma_G)^\perp$, (where the orthogonal
complement is taken within $II^{1,25}$)   is the fixed point sublattice.

The $G$ action on $II^{1,25}$ defines $G$ as a subgroup of $\Aut(II^{1,25})$.
On the other hand, it is known (see \cite{Conway} Chapter 27) that
$\Aut(II^{1,25}) \cong \Aut^+(II^{1,25}) \times \IZ_2$ where $\Aut^+(II^{1,25})$
is the subgroup of transformations preserving the future lightcone and $\IZ_2$ is
generated by the transformation $x \to - x$. This latter inversion symmetry
has no nonzero fixed vectors, but $\Pi$ is a nontrivial lattice, and hence
$G \subset \Aut^+(II^{1,25})$.

Now, choose any root vector $r$ and let $H_r = \{ v\in II^{1,25} \vert v \cdot r =0 \}$.
We claim that $\Pi$ cannot sit inside $H_r$. Otherwise it would follow that
$r\in \iota(\Gamma_G)$. However,   $\iota(\Gamma_G)$ has no root vectors, since, by hypotheses of
our theorem $\Gamma_G$ had no root vectors. The complement of the hyperplanes
in the vector space $V = II^{1,25}\otimes \IR$ is a disjoint union of connected components
called \emph{chambers}. It   follows that
there must be some vector $u \in \Pi$ in  one of the chambers.

Now, (again see \cite{Conway}, Chapter 27)
\be\label{eq:AutPls}
\Aut^+(II^{1,25}) \cong W \rtimes Co_{\infty}
\ee
where $W$ is the Weyl group, generated by reflections in the root vectors, and
$Co_{\infty} \cong \Lambda \rtimes Co_0$ is the affine automorphism group of the
Leech lattice. Amongst the chambers, there is a distinguished
one,  the fundamental Weyl chamber $\CC_0$,
whose walls are the hyperplanes $H_r$ for the set of \emph{Leech roots}. These
are the roots that satisfy $\textbf{n}\cdot r =-1$. (The Leech roots are in 1-1 correspondence with
vectors in the Leech lattice.)  Moreover, it is known that $\textbf{n}\in \CC_0$
and  $Co_\infty$ acts trivially on $\CC_0$. Since $W$ acts simply transitively on the set of chambers, we can choose our embedding of $\Gamma_G$ in $II^{1,25}$ such that $\Pi$ contains a vector $u$ in the fundamental Weyl chamber. Because $\CC_0\cap v(\CC_0)=\phi$ for any non-trivial $v\in W$, it follows that $G\subset Co_{\infty}$. Since $Co_{\infty}$ fixes the null vector $\textbf{n}$, it follows that $G\subset Co_{\infty}$ also fixes $\textbf{n}$, and therefore, $\textbf{n}\in \Pi$. But this implies $\Pi^\perp \subset \textbf{n}^\perp$.
But $\Pi^\perp=\iota(\Gamma_G)$. Furthermore, since $\Gamma_G$ has definite
signature $\textbf{n}\notin \Gamma_G$, so it follows that the projection
$\pi: \textbf{n}^\perp \to \textbf{n}^\perp/\textbf{n}\IZ$ has no kernel when restricted to $\iota(\Gamma_G)$.

Thus, we have obtained an embedded copy of $\Gamma_G$, together with its $G$-action,
as a sublattice of the Leech lattice. We can now extend the action of
 $G$ to the rest of $\Lambda$ so that $\pi(\iota(\Gamma_G))$ using our Lemma 1 above.
Finally, we note that
\begin{equation}
    \text{rk }(\Lambda^G) = 24 - \rk(\Gamma_G) = 2 + \rk(\Gamma^G) \geq 5.
\end{equation}
which completes the proof.

\subsection{Condition 2 Implies Condition 1}

Now suppose we have a subgroup $G\subset Co_0$ whose action on $\Lambda$ leaves an invariant sublattice $\Lambda^G\subset \Lambda$ with
rank $R:=\rk(\Lambda^G)\geq 5$ and discriminant form $q$.
We use Nikulin's embedding theorem 1.12.2b
to  establish the existence of the embedding,
\begin{equation}\label{eq:LamGmin1}
   \iota:  \Lambda_G(-1) \hookrightarrow \Gamma,
\end{equation}
Recall that Nikulin's theorem states that
the only obstruction to the existence of the embedding is the existence of
an even integral lattice with the appropriate signature and discriminant
form to be the orthogonal complement of the embedded copy of $\Lambda_G(-1)$.
If there were an embedding \eqref{eq:LamGmin1} then its orthogonal
complement within $\Gamma$ would have invariants
\begin{equation}
    \{(3,19-(24-R),-q \}=\{(3,R-5,-q \}
\end{equation}
But condition 2b posits the existence of such a lattice, therefore there is an embedding.

Now, we can use Lemma 1 and Remark 1 to  extend the $G$-action from $\iota(\Lambda_G(-1))$ to all of
 $\Gamma$ such that $\iota(\Lambda_G(-1))$ is the coinvariant lattice. Therefore, the sublattice
 of vectors fixed under all of $G$ contains a positive definite $3$-plane. By modifying the group
 using $-1$, if necessary, we can arrange the group is in $O^+(\Gamma)$. $\spadesuit$

\subsection{Alternative Formulations Of Condition 2 Of Lemma 2}

Condition 2 of Lemma 2 does not appear to be a very practical condition.
In this section we give some alternative formulations of Condition 2.

First, we can again use  Nikulin's embedding theorem 1.12.2b to state that the
existence of a suitable even integral lattice is equivalent to the existence of
a primitive embedding
\begin{equation}\label{eq:AltEmbed}
\iota: \Lambda^G(-1) \hookrightarrow II^{R-5,R+3}\cong E_8(-1)\oplus U^{\oplus (R-5)}
\end{equation}
From Nikulin's 1.12.2b we know that such a primitive embedding is equivalent to the existence of an even integral
lattice with complementary invariants $\{(R-5, 3, q)\}$. By switching the sign of the quadratic
form we get an even integral lattice with invariants $\{(3,R-5,-q)\}$.

Next, Nikulin's criterion 1.12.2d gives a \underline{necessary} condition for the existence
of the required embedding \eqref{eq:AltEmbed} as the condition
\footnote{The computation is simply
\be
\ell_+(II^{R-5,R+3}) + \ell_-(II^{R-5,R+3}) - \ell_+(\Lambda^G(-1)) - \ell_-(\Lambda^G(-1))
=  (2R-2) - R = R-2
\ee
}
\be\label{eq:NikNec1}
\begin{split}
R-2
& \geq \ell(\Lambda^G)\\
\end{split}
\ee
It is very useful to define the quantity
\be\label{eq:alpha-def}
\alpha:= R- \ell(\Lambda^G)
\ee
so that \eqref{eq:NikNec1} is just the statement that $\alpha \geq 2$. One of the reasons the quantity
$\alpha$ is so useful is that it is tabulated in the tables of H\"ohn and Mason.  By Nikulin's Corollary
1.12.3 a \underline{sufficient} condition for the desired primitive embedding is the condition
$\alpha > 2$. The application of Nikulin's theorem to the cases $\alpha = 2$ is problematic
because one must then check subsidiary conditions on the $p$-Sylow subgroups of the discriminant
group and these can be difficult or tedious to verify.

\subsection{Finding The Explicit List Of Isometry Groups}

Lemma 2 gives us a way to relate the finite groups that can be subgroups of the hyperk{\"a}hler isometry
group of a \underline{smooth} K3 manifold, to subgroups of the Conway group $Co_0$.
The list of groups that satisfy Lemma 2 constitute the list of finite groups compiled by Xiao \cite{Xiao} and separately by Hashimoto \cite{Hashi}. They are all subgroups of the hyperk{\"a}hler isometry group of
some K3 manifold.
\paragraph{}

In order to produce a useful list we must address the following subtlety.
Suppose $G$ is the hyperk{\"a}hler isometry group of some K3 manifold $(X,g)$.
Then of course any proper subgroup $G_0 \subset G$ acts as a group of hyperk\"ahler
isometries. It might be that $G_0$ appears as the full hyperk\"ahler isometry group
of some other K3 manifold $(X,g')$ with $g' \not= g$, in which case $G_0$ should
also be on our list, or it might be for every other $(X,g')$ such that $G_0$ acts
as a group of hyperk\"ahler isometries in fact $G_0$ turns out to be a proper
subgroup of the full hyperk\"ahler isometry group. In this latter case we do not
want $G_0$ to appear on our list. Once again lattice techniques come to the rescue
because we can rephrase the above question in terms of maximality of subgroups of
$O^+(\Gamma)$ that fix a positive 3-plane $P\subset \Gamma \otimes \IR$.

In the list of groups associated with both the Leech lattice $\Lambda$ and the $K3$ lattice $\Gamma$
we introduce a notion of \underline{maximality} as follows:

\begin{enumerate}

\item  In
the H{\"o}hn-Mason list each of the
fixed point sublattices $\Lambda^G$ (up to conjugacy)  is associated with the \underline{largest} finite group $G$
that fixes it. Therefore, for  any group $G$ that appears on the H{\"o}hn-Mason list,there exists no larger group $H\subset
Co_0$ of which $G$ is a proper subgroup, which fixes the same sublattice $\Lambda^G$ as $G$. Put differently, any group
$H$ that contains $G$ as a proper subgroup must fix a \underline{proper} sublattice of $\Lambda^G$.

\item Hashimoto's list of groups that act holomorphically-symplectically on some K3
surface $(X,J)$ also lists  the genus of the quadratic form ($q_n$) for each $\Gamma^G$.
In fact, as noted on page 33 of \cite{Hashi} (and proved again by us below) the genus
determines the isomorphism class of the fixed point lattice. We can thus compile a
``Hashimoto maximal sub-list'' consisting of groups such that a fixed point lattice
appears just once on Hashimoto's list. Moreover, if two or more groups have fixed point
lattices in the same genus then there is a maximal group (under inclusion) and we just
list the maximal group.  Therefore, for any group $G$ that appears on Hashimoto's maximal sub-list,there exists no larger finite group
$H\subset O^+(\Gamma)$ of which $G$ is a proper subgroup, which fixes the same sublattice $\Gamma^G$ as
$G$. Put differently, any group $H$ that contains $G$ as a proper subgroup must fix a \underline{proper} sublattice of
$\Gamma^G$.

\end{enumerate}

A second subtlety we must dispense with concerns the distinction between fixed sublattices and
fixed linear subspaces of the ambient vector space.
In associating hyperk{\"a}hler isometry groups to the positive 3-planes in $\Gamma^{3,19}\otimes
\mathbb{R}$ that they fix, we must also settle the following question: if we have a group $G$ that is a
proper subgroup of another allowed isometry group $H$, and $H$ has a fixed point sublattice that is not
isomorphic to the fixed point sublattice of $G$, can $H$ and $G$ still fix the same set of positive 3-
planes in $\Gamma^{3,19}\otimes\mathbb{R}$? If this is so, then every positive 3-plane fixed by $G$ is also fixed by $H$, so $G$ should not appear on the list of maximal isometry groups. In the following paragraphs, we will show that this cannot happen.

If two groups $G$ and $H$ that are subgroups of $O^+(\Gamma^{3,19})$ are related as $G\subset H$, then $
\Gamma^H\subset \Gamma^G$. The two lattices may be isomorphic to each other, in which case $G$ and $H$
fix the same set of positive 3-planes in $\Gamma^{3,19}\otimes\mathbb{R}$. If however, $\Gamma^H$ is a
proper sublattice of $\Gamma^G$, then we want to show that in fact, we can always find a positive 3-plane that is fixed by $G$ but not
by $H$.

We are considering the situation where $\Gamma^H$ is not isomorphic to $\Gamma^G$. This implies that we
can find a vector $v\in \Gamma^G$ that is not fixed by the action of the group $H$.
Now, let $P$ be a positive 3-plane in $\Gamma^{3,19}\otimes\IR$ that is fixed by both $G$
and $H$. Also, let $w\in \Gamma^G$ be a vector in the 3-plane $P$.
We can then rotate the plane $P$ by a small angle about the 2-plane generated by $v$ and $w$ to
obtain a new 3-plane $P'$ such that $P'$ also contains non-zero lattice vectors. The resultant 3-plane
$P'$ will be fixed by $G$ but not $H$. Further, we can perform a rotation by an arbitrarily small angle
so that the 3-plane does not cross any null directions. This ensures that the plane $P'$ will also be a
positive 3-plane (Notice that the above 2-plane need not be of definite signature, so that
the rotations we work with are actually boosts, but this does not affect our argument). Thus, whenever
$\Lambda^H$ is a proper sublattice of $\Lambda^G$, both $G$ and $H$ can be realized as maximal
hyperk{\"a}hler isometry groups for two different positive 3-planes.
\paragraph{}

We now prove two lemmas that will help us pin down the list of possible hyperk{\"a}hler isometry groups.

\bigskip
\noindent
\textbf{Lemma 3:} Consider a finite group $G$ that appears on the H{\"o}hn-Mason list as part of the
triple, $(G,\Lambda^G,\Lambda_G)$ (group, fixed point sublattice, orthogonal sublattice) with
\be
R=\text{rk}
(\Lambda^G)\geq 5 \qquad {\rm and} \qquad \alpha=R-\ell(\mathcal{D}(\Lambda^G))\geq 2.
\ee
Then $(G,L:=\iota(\Lambda_G), L^{\perp})$ (where $\iota$ is a primitive embedding, $\iota:\Lambda_G(-1)
\hookrightarrow \Gamma$, and $L^{\perp}$ is its orthogonal sublattice in $\Gamma$) must be a corresponding triple appearing in Hashimoto's maximal sublist.

\bigskip
\noindent
\textbf{Proof:} We divide the proof into two parts:

Case (a):  $R\geq 5$ and $\alpha>2$, then the primitive embedding,
\begin{equation}
   \iota:\Lambda_G(-1)\hookrightarrow \Gamma
\end{equation}
necessarily exists. By Remark 1, $G$ acts trivially on the discriminant group of $\Lambda_G$. Then, by Lemma 1, the induced $G$-action on $\iota(\Lambda_G)$ can be extended (by
identity on the orthogonal complement of $\iota(\Lambda_G)$ in $\Gamma$) to the whole of $
\Gamma$. Therefore, $G\subset O^+(\Gamma)$ and $\Gamma_G\cong \iota(\Lambda_G)$.
Now, we can consider the possibility that there exists a finite group $H\subset O^+(\Gamma)$ such
that $G\subset H$ as a proper subgroup and $\Gamma_H\cong \Gamma_G$. Then, by Lemma 2, this would imply
that $H\subset Co_0$ with $\Lambda_H\cong \Lambda_G$. However, $G$ appears on the H{\"o}hn-Mason list,
the maximality condition of which tells us that this is not possible. Therefore, we conclude that no such
finite group $H$ exists. Thus, $G$ is an element of Hashimoto's maximal sublist. \\

Case (b):  If $R\geq 5$ and $\alpha=2$, then we resort to direct inspection of the H{\"o}hn-Mason and the
Hashimoto lists. Doing so, we can verify that the group $G$ also appears on Hashimoto's maximal sublist
(we call the corresponding coinvariant sublattice $\Gamma_G$). We need to verify that $G$ corresponds to
a triple with an isomorphic coinvariant sublattice on the Hashimoto maximal sublist.\\
By Lemma 2, the primitive embedding $\iota:\Gamma_G\hookrightarrow \Lambda$
exists, and by the maximality condition of Hashimoto's maximal sublist, there exists no $H\subset O^+
(\Gamma^{3,19})$ such that $G\subset H$ as a proper subgroup and $\Gamma_H\cong \Gamma_G$. Since the
H{\"o}hn-Mason list consists of all the isomorphism classes of fixed point sublattices of the Leech
lattice, $L:=\iota(\Gamma_G)$ must be on the H{\"o}hn-Mason list, and it can only correspond to the group
$G$ (and not a larger group H) on the H{\"o}hn-Mason list. Only one lattice has the same genus as on
Hashimoto's list, so in this case the genus fixes the lattice.
Now, by inspection, we see that $L$ is in the same genus as $\Lambda_G$, so $\Lambda_G$ must be
isomorphic to $\iota(\Gamma_G)$. Therefore, $\Lambda_G$ primitively embeds in $\Gamma^{3,19}$, proving
Lemma 3. $\spadesuit$

\bigskip
\bigskip
\noindent
\textbf{Lemma 4:} If a finite group $G$ appears on Hashimoto's maximal sublist, then $G$ must also appear
on the H{\"o}hn-Mason list.

\bigskip
\noindent
\textbf{Proof:} By Lemma 2, $G$ is a subgroup of $Co_0$ such that $\Lambda^G\geq 5$ and $\alpha\geq 2$.
Also, since the H{\"o}hn-Mason list is a complete list of isomorphism classes of fixed point sublattices
of the Leech lattice, $\iota(\Gamma_G)$ must be on this list. The only caveat is that the lattice $\iota(\Gamma_G)$ may be listed as $\Lambda_H$ for some group $H$ that contains $G$ as a proper subgroup.

However, the maximality condition of Hashimoto's maximal sublist implies that no larger group $H\subset Co_0$ that contains $G$ as a proper subgroup, can have $\Lambda_H\cong \Lambda_G$. Therefore, the triple, ($G, \Gamma_G, \Gamma^G)$ must appear on the H{\"o}hn-Mason list as the triple,
\begin{equation}
   (G, \Lambda_G\cong \iota(\Gamma_G), \Lambda^G\cong (\Lambda_G)^{\perp})
\end{equation}
 This proves Lemma 4. $\spadesuit$

\bigskip
\bigskip
\noindent
\textbf{Main Theorem:} The same finite groups appear in Hashimoto's maximal sublist and the entries in
the H{\"o}hn-Mason list that satisfy $R\geq 5$ and $\alpha\geq 2$. The corresponding coinvariant
sublattices $\Gamma_G$ and $\Lambda_G$ are isomorphic.

\bigskip
\noindent
\textbf{Proof:} Follows directly from Lemma 3 and Lemma 4.
\bigskip

This list of groups is the list of hyperk{\"a}hler isometry groups of K3 manifolds. The full list is
given in the appendix, and includes 40 out of the 81 entries in the full Xiao-Hashimoto list.

\bigskip
\noindent
\textbf{Remarks}:

\begin{enumerate}

\item  By inspection of the H\"ohn-Mason list we find there are forty
(isomorphism classes of) finite groups that can and do act as full hyperk\"ahler isometry
groups of K3 manfiolds. The list inclusdes the eleven maximal  Mukai groups. The full list is given in the Appendix.

\item   Here are some examples of finite groups that have a symplectic action on a K3 surface,
but are not full hyperk{\"a}hler isometry group of any K3 manifold $(X,g)$. Notice that groups like $
\mathbb{Z}_5$ and $\mathbb{Z}_8$ appear on Xiao's list of groups and
hence have symplectic actions on a K3 surface. However, these groups are not on H{\"o}hn and Mason's list. So, any K3 surface that has $\mathbb{Z}_5$ as a subgroup of its hyperk{\"a}hler isometry group, will have it as a proper subgroup. From a comparison of quadratic forms, we see that $\mathbb{Z}_5$ has the same fixed point sublattice as $D_{10}$, so any group that has $\mathbb{Z}_5$ as a subgroup of its hyperk{\"a}hler isometry group will also have at least $D_{10}$ isometry. Similarly, $\mathbb{Z}_8$ has the same fixed point sublatice as the semidihedral group of order 16. \\
In fact, the list of hyperk{\"a}hler isometry groups of K3 surfaces can be seen as
the intersection of the groups appearing on the Hashimoto-Xiao list, and the groups appearing on H{\"o}hn and Mason's list with the condition that $\text{rk }(\Lambda^G)\geq 5$ and $\alpha \geq 2$.

\item The largest group of hyperk{\"a}hler isometries on our list is the group $M_{20}\cong 2^4:A_5$, of order 960. This is the symplectic automorphism group of a K3 surface in the family of K3 surfaces known as the Dwork pencil. The general member of this family is the following quartic in $\mathbb{P}^3$,
\begin{equation}
  X^4+Y^4+Z^4+W^4+4\lambda XYWZ
\end{equation}
and its solution is necessarily a Kummer surface \cite{Bini}. For $\lambda=0$, we have the Fermat quartic, which has symplectic automorphism group $F_{384}\cong 4^2:S_4$. For $\lambda=-3$, we get a quartic K3 with symplectic automorphism group $M_{20}$. \\
It is interesting to note that the largest possible automorphism group of a K3 sigma model that commutes with $(4,4)$ supersymmetry is isomorphic to $2^8:M_{20}$. The sigma model with this symmetry was studied in \cite{GTVW, Harvey:2020jvu}. The theory can be described as a $\mathbb{Z}_2$ orbifold of a $D_4$ torus SCFT. Geometrically, the Kummer surface corresponding to the $T_{D_4}^4/\mathbb{Z}_2$ orbifold has the symmetry group $T_{192}\cong 2^4:A_4$ of order 192 (see Section 4.1 of \cite{GTVW}). It would be interesting to see if this SCFT can alternately be described as a sigma model on the quartic K3 surface from the Dwork pencil family. This would require the geometric symmetries of some description of the model to generate the $M_{20}$ group. As verified in \cite{Harvey:2020jvu}, the geometric symmetries of the orbifold theory of the $D_4$ torus SCFT do not generate the $M_{20}$ group.
\end{enumerate}

\section{Implications For Physics}\label{sec:PhysicalImplications}

\subsection{Classical vs. Quantum Symmetries}

It is interesting to compare our result with that of \cite{Gaberdiel:2011fg}.
In this paper we are using the same techniques that were used in
\cite{Gaberdiel:2011fg} but the question we address is slightly different.

Let us recall a few basic facts about supersymmetric sigma models on K3
manifolds. (See, for example,  \cite{Aspinwall:1994rg,Asp,Nahm:1999ps}.)
The data needed to specify the sigma model consists of a
hyperk{\"a}hler metric, and a flat $B$-field. The hyperk{\"a}hler metric is specified by choosing a
positive 3-plane $\Sigma\subset H^2(K3;\mathbb{R})$ and a volume factor $V\in\mathbb{R}_+$. The $B$-field
can be identified with an element of $H^2(K3;\IR)$. Now, we consider the triple $(\Sigma,V,B)$ and use it to construct a positive definite 4-plane inside the full cohomology space $H^*(K3;\IR)$  as follows. Define
\begin{equation}
   \begin{split}
      &\xi:\Sigma\to H^*(K3,\mathbb{R}), \quad \xi(\omega):=\omega-\langle B,\omega \rangle v \\
      &\xi_4:=1+B+(\frac{B^2}{2}-V)v
   \end{split}
\end{equation}
where $v$ is a generator of $H^4(K3;\mathbb{Z})$ while $1\in H^0(K3;\mathbb{R})$. We have $\langle 1,v \rangle=1$
so $1,v$   span the a copy of the hyperbolic plane $U$. We let   $\Pi\subset H^*(K3;\mathbb{R})$ be the positive 4-plane spanned by
$\xi_4$ and the vectors $\xi(\omega)$ for $\omega \in \Sigma$. This is the four-plane of the Aspinwall-Morrison theorem.
The quadratic form on $v = (v_0, v_2, v_4) \in H^*(K3;\IZ)$ (extended linearly to $H^*(K3;\IR)$) is:
\be
\langle v, v\rangle := \int_{K3}[ v_2^2 - 2 v_0 v_4 ]
\ee
and we can use this to check that   $\langle \xi_4,\xi_4 \rangle=2V\in \mathbb{R}_+$ is positive.
 Finally, we need to show that $\Pi$ is positive definite. But this follows easily
 since when restricted to $\Sigma$ the map $\xi$ is an isometry and moreover
 $\langle \xi_4, \xi(\omega) \rangle = 0 $ for every $\omega \in \Sigma$.
%\begin{equation}
%   w=\sum_{i=1}^3 x_i\xi(\omega_i)+x_4\xi_4
%\end{equation}
%has positive definite norm.
%
%Using the inner product on $H^*(K3;\mathbb{Z})$ we
%\begin{equation}
%   \langle v^{(1)}, v^{(2)}\rangle=\int_{K3}[v_2^{(1)}v_2^{(2)}-v_0^{(1)}v_4^{(2)}-v_4^{(1)}v_0^{(2)}]
%\end{equation}
%
%It is straightforward to check that
%\begin{equation}
%   \langle \xi(\omega_i),\xi(\omega_j)\rangle=\langle \omega_i,\omega_j\rangle=0 \text{  for }i\neq j
%\end{equation}
%which vanishes by virtue of being orthogonal generators of the positive 3-plane $\Sigma\subset H^2(K3;\mathbb{R})$. Further,
%\begin{equation}
%   \langle \xi(\omega_i),\xi_4\rangle=\langle B,\omega_i\rangle-\langle B,\omega_i\rangle\langle 1,v\rangle=0
%\end{equation}
%Therefore, any vector $w\in \text{span}_{\mathbb{R}}\{\xi(\omega_i),\xi_4\}$ is positive definite (for $i=1,2,3$).

We can now compare classical and quantum symmetries. For a sigma model
 $\sigma(X,g,B)$ a diffeomorphism  $f$ will be a classical symmetry
 if it preserves the Lagrangian. So, $f$ should be an isometry such that $f^*B = B$.
 Therefore $f^*$ preserves the space $\Pi$. Note that, for all hyperkahler
 isometries there will exist a space of $B$ fields for which those isometries
 are classical symmetries. (The $B$ field must simply be in the space
 fixed by $f$. By averaging over $G$ we know this subspace has positive dimension.)
 By contrast automorphisms of the K3 lattice that preserve $\Pi$ might not
 be induced by a hyperkahler isometry. Such symmetries are quantum symmetries
 of the sigma model.  They do not leave the sigma model action invariant and consequently are not
obviously symmetries of the quantum theory.
 Let us compare the conditions for classical and quantum symmetries:

Suppose that $G\subset Co_0$ and we consider a fixed point sublattice in the Leech lattice $\Lambda^G$
with invariants $(R, 0, q)$, where $R\geq 4$. Then $\CI(\Lambda_G(-1)) = \{ 0, 24 - R, -q\}$. The
necessary condition of Nikulin for the embedding $\Lambda_G(-1) \hookrightarrow H^*(K3;\IZ)$ is simply
\be
R \geq \ell(\Lambda^G)
\ee
which is always automatically true, by equation \eqref{eq:RankEll}. By
contrast the same necessary condition for the embedding into $\Gamma^{3,19}$ to exist (which is what we work with) is the requirement $\alpha \geq 2$, and is a nontrivial restriction on $\Lambda^G$ containing more information than $R \geq 5$.

According to Nikulin the only obstruction to the existence of the embedding, $\Lambda_G(-1) \hookrightarrow H^*(K3;\IZ)$, is the existence of an even integral lattice with invariants
\be
\CI( \iota(\Lambda_G(-1))^\perp ) = (4, 20 - (24-R), -q )
\ee
Now, we can prove that there always exists the following primitive embedding,
\begin{equation}
   \Lambda^G\hookrightarrow II^{R+4,R-4}
\end{equation}
as a result of another of Nikulin's theorems: an even lattice with signature $(p,q)$ necessarily embeds
primitively in an even unimodular lattice with signature $(r,s)$ if $(p+q)\leq (r+s)/2$.\\
The orthogonal complement of $\Lambda^G$ in this embedding is an even lattice with the invariants $\{ 4, R-4, -q \}$, whose existence in turn proves the existence of the embedding, $\Lambda_G(-1) \hookrightarrow H^*(K3;\IR)$.
\paragraph{}
Thus, in the quantum case $R \geq 4$ is necessary and
sufficient. But in the classical case $R\geq 5$ is necessary, but not sufficient.
Rather it is the pair of conditions $R\geq 5$ and $\alpha \geq 2$  that is necessary
and sufficient.

\subsection{Components Of The Low Energy Gauge Group}

We can use a K3 manifold $(X,g)$ to compactify both $M$-theory and
$IIA$-theory to $7$ and $6$ dimensions, respectively.
In these cases the symmetry groups $G$ have an important
physical interpretation related to the components of the
gauge group of the effective theory in the noncompact
directions. In either case, since we assume $X$ to be
smooth the connected component of the gauge group $H_{\rm gauge}$ is
isomorphic to $U(1)^{22}$ and there is an exact sequence
\begin{equation}
    1\to U(1)^{22}\to H_{\rm gauge} \to \pi_0(H_{\rm gauge})\to 1
\end{equation}
The $U(1)^{22}$ gauge potential arises from the harmonic modes
of the 3-form potential in the respective supergravity theories.
We are concerned here with the group of components.

In the case of $M$-theory we can compactify on the
11-dimensional manifold $\CX = \IM^{1,6} \times X$.
We should choose the Riemannian metric $g$ to be
hyperk\"ahler. The $C$-field must be flat and hence
must be trivial. In Kaluza-Klein theory isometry
groups of the compactification space become gauge groups
of the lower-dimensional theory so
$\pi_0(H_{\rm gauge})$ should be identified with
the isometry group of $(X,g)$. In this paper we
have only been able to say something about the
subgroup $ {\rm HKIsom}(X,g)$ of hyperk\"ahler
isometries. This subgroup does have a nice
physical meaning: it is the group of components
of the gauge group that commute with the spacetime supercharges.
\footnote{The hyperk\"ahler isometries will preserve the
covariantly constant spinors and hence commute with the
spacetime supersymmetries.}

\textbf{Remark:} The finite groups studied here are all subgroups of $M_{23}$. It
is interesting to ask if this continues to be the case when generalizing to the
full isometry groups of $(X,g)$. Note that discrete symmetries, both global and
gauged, have been explored in explaining the hierarchy of masses in standard model
phenomenology \cite{LeurerSeiberg}. Here, we encounter examples where
the discrete group is an interesting finite group. It would be interesting
to see if the kind of exotic symmetry groups discussed here could be used
in string phenomenology to explain mass hierarchies.

\subsection{Application To Heterotic/M-theory Duality}

In \cite{Harvey:2017xdt} a distinguished set of Narain compactifications
of the heterotic string called CSS (Conway Subgroup Symmetric) compactifications
 were examined. These are particularly nice
compactifications exhibiting large nonabelian gauge symmetry groups
with Abelian connected component. The essential idea is to choose
a subgroup $G \subset Co_0$ and then consider   $\mathfrak{F}_L := \Lambda^G$.
When there is an isometric embedding $\mathfrak{F}_R$ of this lattice into the $E_8$ lattice we
can form a Narain lattice using the orthogonal lattices $\Lambda_G$ on
the left and $\mathfrak{F}_R^\perp$ on the right. The result is a decompactification of the heterotic string
on the Leech torus times the E8 torus to $d+2$ dimensional Minkowski
space, where $d$ is the rank of $\mathfrak{F}_L$. It is of interest to
know how many of the discrete nonabelian gauge group compactifications
can be obtained using this simple construction.

For the CSS compactification procedure with  $\rk(\Lambda^G)=5$
one requires an embedding into the $E_8$ lattice. The obstruction
to this is just
\be
3 \geq \ell(\CD(\Lambda^G))
\ee
which is equivalent to $\alpha = 5 - \ell \geq 2$. Thus,
CSS compactifications are dual to all the M-theory compactifications
with $\rk(\Lambda^G) = 5$. The discussion of this paper makes clear that
many M-theory compactifications, namely those corresponding to $G$ with
$\rk(\Lambda^G)>5$ will in general not have heterotic duals given by the
CSS procedure. Thus, the CSS procedure does not capture all the Narain
compactifications with interesting disconnected gauge groups. A possible future
direction is therefore to understand if the CSS construction can be generalized to
include all possible discrete gauge symmetry groups of M-theory on a K3 background.

\section{Future Directions}\label{sec:FutureDirections}

We mention two avenues for further investigation.

First, in this paper we have limited attention to smooth K3 manifolds. The notion of
a hyperk\"ahler metric makes sense as well for K3 surfaces that have ADE singularities,
and it would be nice to extend the classification to include these.
 For such singular K3 manifolds, we must relax the condition that $\Gamma_G$ contains no root vectors.
In Lemma 2, instead of embedding into the Leech lattice we should find a primitive embedding into one
of the 23 Niemeier lattices with root vectors. Finite subgroups of $O^+(\Gamma)$ can then be identified
with subgroups of the automorphism groups of these lattices. In order to carry this out we would need
an analog of the H\"ohn-Mason list for all the Niemeier lattices. To our knowledge it is not available.

Second, as we mentioned in the introduction the full isometry group $\text{Isom}(X,g)$
of a K3 manifold is certainly of interest. Now the hyperk\"ahler isometry group
$\text{HKIsom}(X,g)$ is  a normal subgroup of the full isometry group,
and sits inside the group extension,
\begin{equation}
   1\to \text{HKIsom}(X,g)\to \text{Isom}(X,g)\to Q\to 1
\end{equation}
where $Q$ is a finite subgroup of $O(3)$. In particular, $\text{Isom}(X,g)$
will always be a finite group. Moreover, the finite subgroups of $O(3)$
have a well-known  ADE classification. Combining this and investiging the
extension problem one could extend our classification of $\text{HKIsom}(X,g)$
to classify all \underline{potential} K3 isometry groups. It would be
even more interesting to determine the full list of hyperk\"ahler isometry
groups that are in fact realized by some K3 manifold.

Finally (we owe this question to D.~Huybrechts) it would be interesting to
see whether these results can be derived using the twistor formulation of
hyperk\"ahler geometry.

\appendix

\section{Some Relevant Definitions And Results From The Theory Of Lattices}\label{app:LatticeResults}

In this section, we collect some results and theorems that we used in this work.

\subsection{Lattices And Discriminant Groups}

The data of an integer lattice $L$ comes equipped with a nondegenerate, integral,
symmetric bilinear form $Q:L\times L\to \mathbb{Z}$.
We can define its dual lattice $L^{\vee}$ as the set of $\mathbb{Z}$-linear maps
from $L$ to $\mathbb{Z}$. This inherits the bilinear form $Q:L^{\vee}\times L^{\vee}\to \mathbb{Q}$.
For an integer lattice, we have $L\subset L^{\vee}$. The discriminant group is defined to be the
finite Abelian group
\be
\CD(L):= L^\vee/L ~ .
\ee
When $L$ is even the discriminant group inherits a bilinear form valued in $\IQ/\IZ$
together with a quadratic refinement $q:\CD(L) \to \mathbb{Q}/\mathbb{Z}$.
We let $\ell(L)$ denote the minimal number of generators of the group $\CD(L)$.
Note that
\be\label{eq:RankEll}
\ell(L) \leq \rk(L)
\ee
This follows because any basis for $L^\vee$ will descend to a set of generators of $\CD(L)$.

Let $\ell_+(L)$ and $\ell_-(L)$ denote the dimensions of the maximal positive and negative definite
subspaces of $L\otimes \IR$. We define the \emph{invariants of $L$} to be the triple
\be\label{eq:InvtsDef}
\CI(L):= \{  \ell_+(L) , \ell_-(L), q \}
\ee
An important result in lattice theory (see, for example \cite{Nik} Theorem 1.10.2) states that
an even integral lattice with invariants $\{r,s,q\}$ (where $q$ is a quadratic function
$q: \CD \to \IQ/\IZ$ on a finite Abelian group $\CD$) exists if
\begin{enumerate}

\item  $(r-s)=$ sign $q$ (mod 8) and,

\item  $r\geq 0, s\geq 0, r+s> \ell(\CD)$

\end{enumerate}
Here, the notation signature (mod 8) refers to the Arf invariant of a finite quadratic form.
When $r>0$ and $s>0$ and the discriminant group is the trivial group there is a unique
such even unimodular lattice, up to isomorphism, and we will denote it as $II^{r,s}$.
For $r=s=1$ we also use the notation $II^{1,1} = U$.

For any lattice $L$ that is acted on by a group $G$ (acting as a lattice automorphism),
we use $L^G$ to denote the sublattice on which $G$ acts trivially. The
\emph{coinvariant sublattice},
defined to be the orthogonal complement of $L^G$ within $L$, will sometimes be denoted as $L_G$:
\be
L_G := (L^G)^\perp ~ .
\ee

\subsection{Lattice embeddings}

We now describe some of the results of Nikulin \cite{Nik} that we used in our arguments. They are conditions for the existence and uniqueness of lattices with particular invariants, and primitive embeddings of even lattices into even unimodular lattices.

If an even integral lattice $L$ with invariants $\{ r,s,q \}$ can be primitively embedded into an
even unimodular lattice $\Gamma$ with signature $(\ell_+, \ell_- )$ then the orthogonal complement
$L^\perp$ inside $\Gamma$ would have invariants $\{ \ell_+ - r, \ell_- - s ,-q\}$. It therefore follows that
a necessary condition for an embedding of $L$ into $\Gamma$ is the existence of an
even integral lattice with such invariants. (In particular, rather trivially, $\ell_+ \geq r$ and $\ell_- \geq s$.)
A surprising and nontrivial
result is Nikulin's result Theorem 1.12.2(b) which states that this is the \underline{only}
obstruction to the existence of a primitive embedding. So the condition is necessary and sufficient.
We use this result several times in the paper.

Nikulin's result 1.12.3 gives a simple sufficient condition for the existence of an embedding.
We must have $\ell_+ -r \geq 0$, $\ell_- - s \geq 0$ and
\be\label{eq:NikIneq}
(\ell_+ + \ell_-) - (r + s) > \ell(L)
\ee
It also turns out that there is a necessary condition
\be\label{eq:NikNec}
(\ell_+ + \ell_-) - (r + s) \geq  \ell(L)
\ee
When this inequality is an equality there are further necessary conditions
for the existence of an embedding. This involves conditions on the $p$-Sylow
subgroups of the discriminant group that can be difficult to check. Fortunately,
the arguments in the present paper do not rely on these more difficult conditions.
Nevertheless, for completeness, we recall Nikulin's result here:

\bigskip
\bigskip

\emph{Nikulin, Theorem 1.12.2d}:
There exists a primitive embedding of an even lattice with invariants ($r_0,s_0,q)$ into another even, unimodular lattice with invariants $(r,s)$ if the following conditions are satisfied:

\begin{enumerate}

\item   $(r+s)-(r_0+s_0)\geq \ell(\CD_q)$

For all primes $p$ that divide the order of the group $\CD_q$, we can define its p-Sylow subgroups $\CD_{q_p}$. The notation $q_p$ means that we can decompose the discriminant form on $\CD_q$ into the direct sum,
\begin{equation}
    q_p=\oplus_p q_p
\end{equation}
where $q_p$ is the restriction of the discriminant form to the p-Sylow subgroup. Note that the bound of condition (1) implies that
\be\label{eq:p-bound}
(r+s)-(r_0+s_0)\geq \ell(\CD_{q_p})
\ee
for all $p$.

\item  For all odd primes $p$ for which the bound \eqref{eq:p-bound} is saturated,
i.e. such that $(r+s)-(r_0+s_0)=\ell(\CD_{q_p})$, we have the condition,
\begin{equation}
    |\CD_q|=\pm \text{discr }(K(q_p))(\text{mod }((\mathbb{Z}_p^*)^2))
\end{equation}
Here $K(q_p)$ is a $p$-adic lattice of rank $\ell(\CD_{q_p})$ whose discriminant form is isomorphic to $q_p$.
It turns out that for odd primes $p$ this lattice is unique. Also,  $\text{discr }S$ is a determinant of the Gram matrix of the Lattice $S$.

\item  If the even prime $p=2$ saturates the bound \eqref{eq:p-bound}, and $q_2\neq q_{\theta}^{(2)}(2)\oplus q'_2$, i.e., the discriminant form does not split off an $su(2)$ factor, then,
\begin{equation}
    |\CD_q|=\pm \text{discr }(K(q_2))(\text{mod }(\mathbb{Z}_2^*)^2)
\end{equation}
(For $p=2$ the lattice $K(q_2)$ is not unique, but we will not need to go into this subtlety for our application.)

\end{enumerate}

Condition (2), as described, is difficult to check explicitly, but if, for
 all odd primes $p$ that divide the order of the discriminant group, we have the strict inequality
\begin{equation}\label{eq:OddPrimeSuff}
r_0+s_0>\ell(\CD_{q_p})
\end{equation}
then condition (2) holds. This easier, sufficient condition, is what we use in our proof.
Similarly, condition (3) is difficult in general but an easier condition is:

(3b) If the even prime $p=2$ satisfies $(r+s)-(r_0+s_0)=\ell(\CD_{q_2})$, then the discriminant form must split off a factor corresponding to the discriminant form on the $\CD_1$ root lattice. The discriminant group of the $\CD_1\cong su(2)$ root lattice is $\mathbb{Z}_2$, a group generated by a single element.

\subsection{The Leech Lattice And Its Automorphism Group}

This section reviews some essential material from Chapter 26 and 27 of \cite{Conway}.

\paragraph{}

 The study of the automorphisms of $II^{1,25}$ starts by looking at its root vectors. These are vectors $r\in II^{1,25}$ such that $\langle r,r\rangle=-2$. We define $R_r$ to be the hyperplane in $II^{1,25}\otimes \mathbb{R}$ perpendicular to the root $r$. Reflections of the lattice vectors about these hyperplanes generate a group  $W$ called the Weyl group of the Leech lattice. It turns out to be a normal subgroup of the automorphism group of the lattice.  The hyperplanes divide the vector space  into chambers $\mathcal{C}$, known as Weyl chambers,  with the hyperplanes forming the walls of the chambers. The Weyl group acts transitively on the Weyl chambers. The roots corresponding to the walls of any one of these chambers can be taken as a set of generators for $W$. They form a set of fundamental roots. Now, there is a distinguished set of fundamental roots that we can use for convenience. Consider the roots $s$ that have the additional property that $\langle s,\textbf{n}\rangle=-1$, where $\textbf{n} = (70;0,1,2,\cdots,24)$ is the characteristic vector of the lattice, a distinguished null vector. They form a set of fundamental roots called the Leech roots. There is a one-to-one correspondence between these roots and points on the Leech lattice. The hyperplanes $R_s$ corresponding to these roots form the walls of a distinguished chamber $\mathcal{C}_0$ called the fundamental chamber. It is the chamber that contains the characteristic vector $\textbf{n}$. \\
Next, the fundamental roots are used to define the Coxeter diagram of the lattice $II^{1,25}$. To draw the Coxeter diagram, we need the following relations, \\
(0) For any root $r$, we have $R_r^2=1$. \\
(1) For each fundamental root $s_i$, we draw a node in the diagram. \\
(2) If the hyperplanes correspond to two roots $r$ and $s$ that have the relation, $(R_rR_s)^2=1$ when acting on a generic lattice element, the two nodes are unjoined. \\
(3) If they satisfy the relation, $(R_rR_s)^3=1$, then the two nodes are joined. \\
(4) Both of these relations come from hyperplanes that intersect. If two hyperplanes do not intersect, we join them with a bold line for parallel hyperplanes, and a dotted line for divergent hyperplanes. \\
For the lattice $II^{1,25}$, there are an infinite number of nodes. The group of automorphisms of the Coxeter diagram is called $Co_{\infty}$. It is an infinite group that is so named because it is abstractly isomorphic to the full automorphism group of the Leech lattice, including translations. It acts transitively on the fundamental roots. The full automorphism group of $II^{1,25}$ has the structure of a direct product, $O(II^{1,25})\cong O^+(II^{1,25})\times \mathbb{Z}_2$, where, $O^+(II^{1,25})\cong W\rtimes Co_{\infty}$ is the autochronous group.

\subsection{The Genus Of A Quadratic Form}

Our proof of the main theorem makes use of the notion of the genus of a
lattice. Therefore we briefly recall here a few relevant facts from the
classification theory of integral quadratic forms. See  \cite{Cassels,Conway} for a systematic treatment. \\
We define a binary quadratic form using the following equation,
\begin{equation}
    f=ax^2+2bxy+cy^2=
    \begin{pmatrix}
       &x &y
    \end{pmatrix}
    \begin{pmatrix}
       &a &b \\
       &b &c
    \end{pmatrix}
    \begin{pmatrix}
       &x \\
       &y
    \end{pmatrix}
    =X^TAX
\end{equation}
where $a,b,c\in\mathbb{Z}$, and we can generalize this from binary to n-ary integral quadratic forms in a straightforward way. Two quadratic forms $f$ and $g$ (with associated matrices $A$ and $B$) are said to be integrally equivalent when the matrices are related as,
\begin{equation}
    B=M^TAM
\end{equation}
where the matrix $M$ has integer entries and $\text{det }(M)=\pm 1$. Using this notion of equivalence of quadratic forms, we can formulate a classification scheme for quadratic forms. \\
Two quadratic forms $f$ and $g$ are said to be in the same genus if and only if,
\begin{equation}
   f\oplus
   \begin{pmatrix}
      &0 &1 \\
      &1 &0
   \end{pmatrix}
   \text{ and }
   g\oplus
   \begin{pmatrix}
      &0 &1 \\
      &1 &0
   \end{pmatrix}
   \text{ are integrally equivalent}
\end{equation}
If two forms are integrally equivalent, then they are in the same genus, but the converse is not true.
\paragraph{}
For any quadratic form, there exists a Jordan decomposition over the p-adic integers. The decomposition
is as follows,
\begin{equation}
    f=f_1\oplus (pf_p\oplus p^2f_{p^2}\oplus \cdots )\oplus(qf_q\oplus q^2f_{q^2}\oplus \cdots)\oplus \cdots
\end{equation}
where $f_q$ is a p-adic unit form (a term we do not define here, merely understanding the above as a
decomposition into a set of standard forms called Jordan constituents that play the role of building
blocks in the classification). For $p=-1$, $f_q$ is simply a positive definite form. We refer the interested reader to Section 7, Chapter 15 of \cite{Conway} for more details. \\
With this Jordan decomposition, we can now define a complete set of invariants to characterize the genus
of a quadratic form, \\
$\text{det }(f_q)$ is the determinant of the Jordan constituent $f_q$ \\
$r=n_q=\text{dim}(f_q)$ is the dimension of the Jordan constituent \\
$d=\epsilon_q=p^{-1}\text{det }(f_q)$ \\
For $p=2$, we will need further invariants (the type $e$ and the oddity $t$ of the form), but we do not
define these here. We can now express the genus using a p-adic symbol, and this is what we will read off
the tables of H{\"o}hn-Mason and Hashimoto. The p-adic symbol for the genus of a quadratic form is a
formal sum over the Jordan constituents,
\begin{equation}
   q^{dr}_{e} \quad \text{for } p\neq 2
\end{equation}
For $p=2$, we need to specify the type and the oddity invariants where applicable. We use the symbol,
\begin{equation}
   q^{dr}_t \text{ where t is the oddity or}\quad q^{dr}_e \text{ where }e\in \{I,II\}
\end{equation}
Finally, we will use the following notation where the oddity $t$ is required to specify the form,
\begin{equation}
    b_t^{dr}=q(L_{r,t,d,I}^{(2)}(b))
\end{equation}
where $L_{r,t,d,e}^{(2)}$ is the unique even unimodular lattice over $\mathbb{Z}_2$ with the invariants,
\begin{equation}
    \begin{split}
        &r=\text{rank }(L^{(2)}) \\
        &d=\pm 1 \\
        &t=\sum_i \epsilon_i \text{ (mod 8}\mathbb{Z}_2) \\
        &e=I \text{ or }II, \quad (I \text{ if the lattice is odd})
    \end{split}
\end{equation}
The genus of a quadratic form is an equivalence relation, which means that two non-isomorphic lattices may have quadratic forms on their discriminant groups that are in the same genus. There are different theorems that decide when there is a unique equivalence class of quadratic forms in a given genus and when there are more than one. One such theorem is as follows,\\
For quadratic forms $f$ having indefinite signature, $|\text{det }f|<128$ and dimension at least 3, there is only one equivalence class of forms in a genus, unless $4^{[n/2]}\text{det }f$ is divisible by $k^{n(n-1)/2}$ for $k=0 \text{ or }1 \text{ (mod}4)$ . \\
We will not require these theorems, as we will only use the result that if two lattices have quadratic forms on their discriminant groups that are not in the same genus, then the two lattices are non-isomorphic.

\section{Some K3 Surfaces With Infinite Holomorphic Symplectic Automorphism Group}\label{app:InfiniteHoloSymp}
\label{sec:Approach1}

In this appendix we discuss K3 surfaces with infinite groups of holomorphic symplectic
automorphisms.  Let $S$ be a positive 2-plane in $H^2(X;\IR)$ representing the
span of the real and imaginary parts of the nowhere zero holomorphic 2-form.
Note that the orthogonal complement to $S$ in $H^2(X;\IR)$ is of signature $(1,19)$
and hence there can in principle be infinite groups of boosts in $O^+(\Gamma)$
acting trivially on $S$. This is the source of infinite
groups of holomorphic symplectic automorphisms.

Indeed, concrete examples can be found using the K3 surfaces with Picard number 20.
These are known as  ``singular K3 surfaces'' (a term we suggest should be deprecated)
or ``Shioda-Inose surfaces.'' They have also been popular with physicists interested
in the interplay between string theory and number theory
\cite{Benjamin:2018mdo,Moore:1998zu,Moore:1998pn,Moore:2004fg,Wendland:2003ma} where
they have been called ``attractive K3 surfaces.''
Shioda and Inose have shown (see \cite{HuyK3} Corollary 2.12, Page 317) that K3 surfaces with Picard number 20 always have an infinite automorphism group (as well as an infinite group of symplectic automorphisms, since this is a finite index subgroup of the automorphism group). The result is an application of the Shioda-Tate formula.
The Shioda-Tate formula relates the Picard number of an elliptically fibered K3 surface to the rank of the Mordell-Weil group. When
this rank is positive translations by the section can define an infinite group of automorphisms.
Shioda and Inose show that an attractive K3 surface is a rational double cover of a Kummer surface $X$ associated with the abelian surface $A$ that is a product of elliptic curves, $A=E_1\times E_2$.
Then the Shioda-Tate formula implies that there always exists a non-torsion section of the map,
\begin{equation}
   X\to E_1/\mathbb{Z}_2 \cong \mathbb{P}^1
\end{equation}
Translation by such a section is an automorphism of infinite order, so the automorphism group of the surface is an infinite group.
Here we content ourselves with a simple explicit example.

 We start with a Kummer surface $X$ associated to an abelian surface $A$ that is a product of two elliptic curves $E_1\times E_2$
 with complex multiplication. To be concrete we consider the Kummer surface
\be
(E_{\tau_1} \times E_{\tau_2})/\IZ_2
\ee
where $E_\tau:= \IC/(\IZ + \tau \IZ)$ and
\be
\tau_1 = \frac{-b + \sqrt{D} }{2a} \qquad \qquad \tau_2 = \frac{b+\sqrt{D}}{2} = -c/\tau_1
\ee
here $D:=b^2 -4ac<0 $, $a,b,c$ are integers and we choose the sign of $\sqrt{D}$ so that $\tau_1$
and $\tau_2$ are in the upper half plane. So $a>0$.
 We will also assume that $a,b,c$ are pairwise relatively prime integers.
Note that
\be
a \tau_1^2 + b \tau_1 +c =0
\ee
from which it follows immediately that
\be
\tau_2 = - \frac{c}{\tau_1} = b + a \tau_1
\ee
It is also worth noting that
\be
\tau_2^2 - b \tau_2 + ac =0
\ee
On the elliptic curves we use coordinates $[z_1]$ and $[z_2]$
with $z_1 \sim z_1 + \IZ + \tau_1 \IZ$ and $z_2 \sim z_2 + \IZ + \tau_2 \IZ$.
The orbifold is simply $(z_1,z_2) \sim (-z_1 , -z_2) $.

It is not difficult to find  conditions on $\alpha, \beta, \gamma, \delta \in \IC$ so that
the transformation
\be
\begin{split}
[z_1] & \to [ \alpha z_1 + \beta z_2 ] \\
[z_2] & \to  [\gamma z_1 + \delta z_2 ] \\
\end{split}
\ee
is well defined. The result is:
\be\label{eq:LegalTmn}
\begin{pmatrix} \alpha \in \IZ + a \tau_1 \IZ& \beta \in \IZ + \tau_1 \IZ  \\  \gamma\in a \IZ + \tau_2 \IZ  & \delta \in \IZ + a \tau_1 \IZ\\
\end{pmatrix}
\ee
Now, note that   $R = \IZ + a \tau_1 \IZ $ is a ring and
\be
R' = a \IZ + \tau_2 \IZ \subset R
\ee
is an ideal. Moreover
\be
R \times (\IZ + \tau_1 \IZ) \subset \IZ + \tau_1 \IZ
\ee
and
\be
(a \IZ + \tau_2 \IZ) \times (\IZ + \tau_1 \IZ) \subset  (\IZ + a \tau_1 \IZ)
\ee
so that the set of matrices in \eqref{eq:LegalTmn} is a monoid under  matrix multiplication.
Therefore, the set of matrices of the above type that is invertible is a group.
Note that the group action descends to the orbifold.

Finally, we want this to be a symplectic automorphism. The symplectic condition
is simply the determinant one condition:
\be
\alpha \delta - \beta \gamma =1
\ee

Now write
\be
\begin{split}
\alpha & := \alpha_1 + \alpha_2 a \tau_1 \\
\beta & := \beta_1 + \beta_2 \tau_1 \\
\gamma & := a \gamma_1 + \gamma_2 \tau_2 \\
\delta & := \delta_1 + \delta_2 a \tau_1 \\
\end{split}
\ee
with $\alpha_i, \beta_i, \gamma_i, \delta_i \in \IZ$ for $i=1,2$.  So we must find
solutions to the Diophantine conditions:
\be
\begin{split}
\alpha_1 \delta_2 + \alpha_2 \delta_1 - b \alpha_2 \delta_2 - \beta_2 \gamma_1 - \beta_1 \gamma_2 & = 0 \\
\alpha_1 \delta_1 -ac \alpha_2 \delta_2 + c \beta_2 \gamma_2 - b\beta_1 \gamma_2 - a \beta_1 \gamma_1 & = 1 \\
\end{split}
\ee
If we put $\delta_1 = 1 $ and $\delta_2 = 0$ we get a pair of equations that we can use to solve for $\alpha_1, \alpha_2$
for any choice of $\beta_i, \gamma_i$. Therefore the group is infinite.
\section{The Possible Hyperk\"ahler Isometries Of K3 Surfaces}\label{app:HyperKList}
\begin{tabular}{|c|c|c|c|c|c|c|c|}
    \hline
         &\# &Rank &Order &G &Index &$\alpha$ &Type  \\
         \hline
         &1 &5 &960 &$2^4:A_5$ &$\#11357$ &2 &$M_{23}^{*}$ \\
         \hline
         &2 &5 &384 &$4^2:S_4$ &$\#18135$ &2 &$M_{23}$ \\
         \hline
         &3 &5 &360 &$A_6$ &$\#118$ &3 &$M_{23}^{*}$ \\
         \hline
         &4 &5 &288 &$A_{4,4}$ &$\#1026$ &2 &$M_{23}^{*}$ \\
         \hline
         &5 &5 &192 &$T_{192}$ &$\#1493$ &2 &$M_{23}^{*}$ \\
         \hline
         &6 &5 &192 &$H_{192}$ &$\#955$ &2 &$M_{23}$ \\
         \hline
         &7 &5 &168 &$L_2(7)$ &$\#42$ &3 &$M_{23}^{*}$ \\
         \hline
         &8 &5 &120 &$S_5$ &$\#34$ &3 &$M_{23}$ \\
         \hline
         &9 &5 &72 &$M_9$ &$\#41$ &2 &$M_{23}$ \\
         \hline
         &10 &5 &72 &$N_{72}\cong 3^2D_8$ &$\#40$ &2 &$M_{23}$ \\
         \hline
         &11 &5 &48 &$T_{48}\cong Q_8\rtimes S_3$ &$\#29$ &2 &$M_{23}^{*}$ \\
         \hline
         \hline
         &12 &6 &192 &$4^2A_4$ &$\#1023$ &2 &$M_{23}^{*}$ \\
         \hline
         &13 &6 &96 &$2^4D_6$ &$\#227$ &2 &$M_{23}$ \\
         \hline
         &14 &6 &72 &$A_{4,3}$ &$\#43$ &3 &$M_{23}^{*}$ \\
         \hline
         &15 &6 &64 &$\Gamma_{25}a_1$ &$\#138$ &2 &$M_{23}$ \\
         \hline
         &16 &6 &60 &$A_5$ &$\#5$ &4 &$M_{23}^{*}$ \\
         \hline
         &17 &6 &48 &$2\times S_4$ &$\#48$ &2 &$M_{23}$ \\
         \hline
         &18 &6 &36 &$3^2Z_4$ &$\#9$ &3 &$M_{23}$ \\
         \hline
         &19 &6 &36 &$S_{3,3}$ &$\#10$ &2 &$M_{23}$ \\
         \hline
         &20 &6 &21 &$F_{21}$ &$\#1$ &3 &$M_{23}^{*}$ \\
         \hline
         &21 &6 &20 &Hol($\mathbb{Z}_5)$ &$\#3$ &3 &$M_{23}$ \\
         \hline
         &22 &6 &16 &$SD_{16}$ &$\#8$ &2 &$M_{23}$ \\
         \hline
         \hline
         &23 &7 &48 &$2^4\times 3$ &$\#50$ &2 &$M_{23}^{*}$ \\
         \hline
         &24 &7 &32 &$2^4Z_2$ &$\#27$ &2 &$M_{23}$ \\
         \hline
         &25 &7 &32 &$Q_8*Q_8$ &$\#49$ &2 &$M_{23}$ \\
         \hline
         &26 &7 &24 &$S_4$ &$\#12$ &4 &$M_{23}$ \\
         \hline
         &27 &7 &8 &$Q_8$ &$\#4$ &2 &$M_{23}$ \\
         \hline
         \hline
         &28 &8 &18 &$A_{3,3}$ &$\#4$ &3 &$M_{23}^{*}$ \\
         \hline
         &29 &8 &16 &$D_8\times 2$ &$\#11$ &2 &$M_{23}$ \\
         \hline
         &30 &8 &12 &$D_{12}$ &$\#4$ &4 &$M_{23}$ \\
         \hline
         &31 &8 &12 &$A_4$ &$\#3$ &4 &$M_{23}^{*}$ \\
         \hline
         &32 &8 &10 &$D_{10}$ &$\#1$ &4 &$M_{23}^{*}$ \\
         \hline
         \hline
         &33 &9 &16 &$2^4$ &$\#14$ &2 &$M_{23}$ \\
         \hline
         &34 &9 &8 &$D_8$ &$\#3$ &4 &$M_{23}$ \\
         \hline
         \hline
         &35 &10 &8 &$2^3$ &$\#5$ &2 &$M_{23}$ \\
         \hline
         &36 &10 &6 &$S_3$ &$\#1$ &5 &$M_{23}$ \\
         \hline
         &37 &10 &4 &4 &$\#1$ &4 &$M_{23}$ \\
         \hline
         \hline
         &38 &12 &4 &$2^2$ &$\#2$ &4 &$M_{23}$ \\
         \hline
         &39 &12 &3 &3 &$\#1$ &6 &$M_{23}^{*}$ \\
         \hline
         \hline
         &40 &16 &2 &2 &$\#1$ &8 &$M_{23}$ \\
         \hline
\end{tabular}
\paragraph{}
\textbf{Notation:} \\
We use the notation of Hashimoto to denote our groups \cite{Hashi}. \\
$2^4$ is used to represent $\mathbb{Z}_2^4$, and in general, exponents represent direct products, positive integers denote cyclic groups of that order. \\
$2^4:A_5$-the semicolon denotes a semidirect product, i.e., the extension splits. $2^4:A_5$ is isomorphic to $M_{20}$, the Mathieu group of a set of 20 elements. \\
$2^4.A_5$ denotes an extension of $A_5$ by $\mathbb{Z}_2^4$, but it is not known whether the extension splits. \\
Hol-denotes the holomorph of a group, i.e., $G\rtimes \text{Aut}(G)$.\\
$SD_{16}$ is the semi-dihedral group of order 16. \\
The group $A_{4,4}$ of order 288 has the structure $A_{4,4}\cong 2^4A_{3,3}$. $A_{3,3}$ is a group of order 18 which is the semidirect product $3^2:2$. It is a generalized dihedral group of the elementary abelian group of order 9. \\
$M_9$ is a Mathieu group, a multiply transitive permutation group on a set of nine elements. \\
$L_2(7)$ or $PSL(2,\mathbb{F}_7)$ is the projective special linear group that acts on a vector space of dimension 2 over the finite field $\mathbb{F}_7$, acting via fractional linear transformations and with determinant equal to 1. \\
$F_{21}$ is the unique nonabelian group of order 21, and also the smallest nonabelian group of odd order. It has the following presentation,
\begin{equation}
    F_{21}=\langle x,y|x^7=y^3=e, xy=yx^2  \rangle
\end{equation}
The last column uses the notation of H{\"o}hn and Mason \cite{HM}. The notation $M_{23}$ implies that $G\subset M_{23}$. The notation $M_{23}^{*}$ implies that $G\subset M_{23}$ and $\widetilde{O^2(G)}=G$. Here $O^2(G)$ is the minimal normal subgroup of $G$ generated by all elements of odd order. It is a subgroup of $ \widetilde{O^2(G)}$, which is a bit technical to describe so the reader should consult \cite{HM} for
the definition. \\
\textbf{Mukai's list:} We see that the eleven groups in the above list for which rk($\Lambda^G)=5$ are precisely the groups that appear on Mukai's list of maximal finite groups that have a symplectic action on a K3 surface. \\
\textbf{Groups of order 192}: Three appear on Xiao's list-$4^2A_4$, $H_{192}\cong 2^4D_{12}$ and $T_{192}\cong(Q_8*Q_8)\rtimes S_3$. Of these, the last two appear on Mukai's list, so they must definitely appear on H{\"o}hn-Mason's list with $\alpha\geq2$. Three distinct groups appear on H{\"o}hn-Mason's list with $\alpha\geq 2$, and they have the following GAP nomenclature. For rk$(\Lambda^G)=5$, $\#955(23)$, $\#1493(23^{*})$. For rk$(\Lambda^G)=6$, $\#1023(23^{*})$. Thus, all three of Xiao's groups must be on our list. \\

\end{document}